\documentclass[twocolumn,pra]{revtex4}
\usepackage{amsfonts}
\usepackage{amsmath}
\usepackage{amssymb}
\usepackage{graphicx}
\usepackage{txfonts}

\newcommand{\aveg}[1]{\left\langle{#1}\right\rangle}
\newcommand{\ket}[1]{\left\vert{#1}\right\rangle}
\newcommand{\bra}[1]{\left\langle{#1}\right\vert}

\begin{document}

\title{Optical and atomic stochastic resonances in the driven dissipative Jaynes-Cummings model}
\author{Qingyang Qiu, Shengdan Tao, Cunjin Liu, Shengguo Guan, Min Xie,}
\affiliation{College of Physics, Electronics, and Communications, Jiangxi Normal University, Nanchang, 330022, China}
\author{Bixuan Fan}
\email{bixuanfan@jxnu.edu.cn}
\affiliation{College of Physics, Electronics, and Communications, Jiangxi Normal University, Nanchang, 330022, China}
\begin{abstract}
In this work we study the stochastic resonance (SR) effect in a driven dissipative Jaynes-Cummings model. The SR effect is systematically studied in the semiclassical and full quantum frameworks and in both cases we find that SRs simultaneously occur for the optical and atomic freedoms. In particular, at zero temperature quantum SR can be induced merely by vacuum fluctuations. The qualitative features of semiclassical SR and quantum SR are similar, but the parameter region of quantum SR are shifted from the semiclassical region due to the widely-used factorization in obtaining semiclassical equations of motion. Our results provide a theoretical basis for experimentally observing and studying the SR phenomenon of the Jaynes-Cummings model in the quantum regime.
\end{abstract}

\maketitle

\section{Introduction}
The Jaynes-Cummings (JC) model \cite{Jaynes,Shore}, as one of most fundamental models in quantum optics, describes the electric-dipole interaction between an atom in the two-level approximation (qubit) and a quantized electromagnetic mode. JC model and its generalized models cover a large part of interactions between fields and natural atoms \cite{Kerckhoff} or artificial matters (such as the superconducting circuit system \cite{Fink} and the quantum dot system \cite{Madsen}). Though the structure of JC model is simple, it has very rich physics, such as collapse-revival phenomenon \cite{Eberly}, optical squeezing \cite{squeezing}, Schr\"{o}dinger cat state \cite{Buzek}, antibunching effect \cite{antibunching}, vacuum Rabi splitting \cite{Tian}, and optical bistabilities \cite{Carmichael,Mavrogordatos}. In particular, JC model is an ideal platform for studying nonlinear dynamics of quantum systems and their interplay with quantum fluctuations \cite{Carmichael,Armen06,Armen11,Mavrogordatos}, which is interesting for fundamental physics research and profoundly relevant with applied quantum science.

Stochastic resonance (SR) \cite{Benzi81,Benzi82,Gammaitoni,Wellens}, as a mechanism exploiting noises to enhance the responses of a nonlinear system to an input weak signal, is a good example of noise induced positive effect on a nonlinear system. SR has shown great potential in detecting faint signals buried in noises in a variety of fields \cite{Gammaitoni,Wellens}. Since 1990s, the study of the SR effect has been extended to the quantum realm, such as the quantum nonlinear oscillator system \cite{Adamyan}, the maser system \cite{Wellens99}, the Dicke model \cite{Witthaut}, the spin-boson model \cite{Grifoni96}, and the quantum optomechanical system \cite{Fan}.

\begin{figure}[pbt]
  \centering
  \includegraphics[width=3.2in]{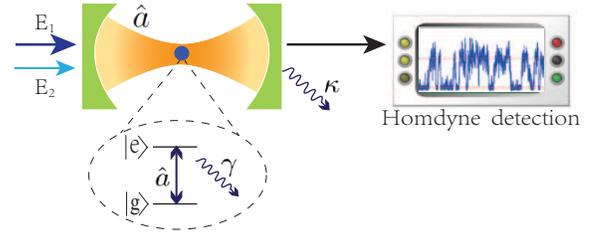}\\
  \caption{Schematic: A two-level atom (a qubit) interacts with a single-model cavity field with two driving fields $E_1$ and $E_2$. $E_1$ is a strong control field resonant with the cavity mode and $E_2$ is a weak signal field slightly detuned from the cavity mode. The output of the cavity is detected by the homodyne detection.}\label{model}
\end{figure}
\begin{figure}[pbt]
  \centering
  \includegraphics[width=3.0in]{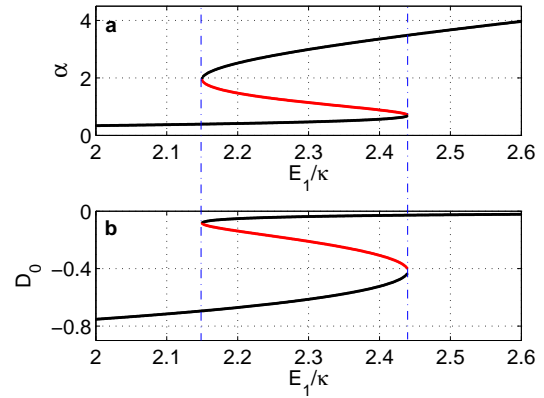}\\
  \caption{Simultaneous bistability in the optical ($\textbf{a}$) and atomic ($\textbf{b}$) freedoms. The parameters are: $\kappa=1$, $\gamma=10\kappa$, $g=6\kappa$, and $E_2=0$.}\label{bistability}
\end{figure}
In this work we study the noise induced stochastic bistable dynamics and the SR effect in a driven dissipative Jaynes-Cumming model in both semiclassical and full quantum frameworks. Through the steady-state solutions and the stability analysis, we find the region for the absorptive bistability, which is shared by the optical and atomic modes. With the activation of noise, the optical and atomic modes have random but simultaneous transitions between two metastable states and the rate of transitions obeys the Kramers law \cite{kramers}. When a subthreshold signal with a suitable frequency is applied, SR occurs in the semiclassical picture, which are characterized via the residence time distribution, synchronization between system responses and the signal, and the SNR resonance peak. In the full quantum description, we use the quantum trajectory method to simulate the system dynamics conditioned on the noisy homodyne currents. We find that at zero temperature vacuum fluctuations can induce spontaneously transitions between metastable states and quantum SR behaviors with a suitable signal. Compared to SR in the semiclassical picture, quantum SR is qualitatively similar but the required parameter region is shifted to larger driving side. This is because that the factorization used in obtaining the semiclassical equations is inappropriate in a strong coupling regime (the cooperation coefficient $C=7.2$). In addition, the system parameters we use are feasible in current experiment conditions, i.e., the single $Cs$ atom QED system \cite{Kerckhoff} and the superconducting circuit system \cite{Fink}, therefore our analysis would lay theoretical basis for experimental observation of SR phenomena in the JC model.

The paper is organized as follows. In Sec. II we introduce the model, find the bistability region for the optics and atom, and show the noise activated stochastic bistable transitions. Then, in Sec. III we show the SR features in the semiclassical description, including the residence time distributions, the synchronization between the input field and the system responses, and the resonance-like effect of the SNR curve.  In Sec. IV, the SR effect in the full quantum mechanical framework is studied and the differences between quantum SR and semiclassical SR are discussed. Finally, we conclude our work in Sec. V.

\begin{figure*}[btp]
  \centering
  \includegraphics[width=6.8in]{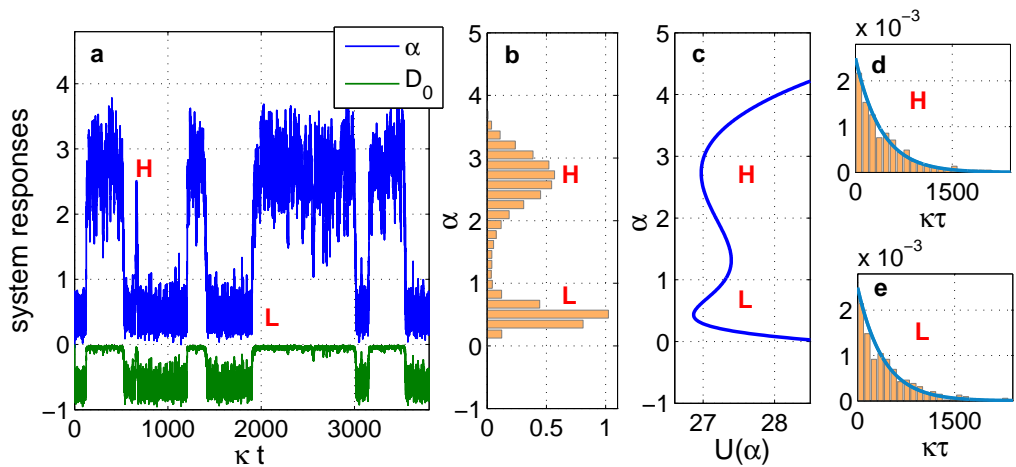}\\
  \caption{$\textbf{a}$ Simultaneous stochastic transitions of the optical field amplitude $\alpha$ and the atomic population inversion $D_0$ in the absence of the signal $E_2$. $H$ ($L$) denotes for the high (low) amplitude state of the field. $\textbf{b}$  The histogram of $\alpha$ using the data (the blue curve) in $\textbf{a}$. $\textbf{c}$ The effective potential function $U(\alpha)$ as a function of $\alpha$, calculated from Eq. (\ref{po}). $\textbf{d}$ and $\textbf{e}$ are the residence time ($\tau$) distributions of the high and low amplitude states of the field for a long evolution time ($T=500000\kappa^{-1}$), respectively. The blue curves show the exponential fit [$1/a$ $\rm{exp(-\tau/a)}$] to the data, with $a=400 \kappa^{-1}$ for both data in $\textbf{d}$ and $\textbf{e}$. The parameters are: $\kappa=1$, $\gamma=10\kappa$, $g=6\kappa$, $D=0.03\kappa$, and $E_1=2.24\kappa$.}\label{bist}
\end{figure*}
\section{Model and Equations}
As shown in Fig. 1, the considered system is a two-level atom (a qubit) interacting with a single-mode cavity field, that is, the well-known JC model. The cavity is driven by two fields: one strong control field $E_1$ with frequency $\omega_{d1}$ and one weak signal field $E_2$ with frequency $\omega_{d2}$.  We assume that the driving field $E_1$ is exactly resonant with the atomic transition frequency and the cavity central frequency. In the rotating frame at the driving frequency $\omega_{d1}$, the Hamiltonian for the described system is given by ($\hbar=1$)
\begin{eqnarray}\label{H2}
\hat{H}_s = g(\hat{\sigma}_+\hat{a} + \hat{a}^\dag\hat{\sigma}_-) -i E_1(\hat{a}-\hat{a}^\dag) -i E_2(\hat{a}e^{i\delta t} -\hat{a}^\dag e^{-i\delta t})
\end{eqnarray}
where $\delta=\omega_{d2}-\omega_{d1}$. $g$ is the atom-field interaction coefficient, $\hat{a}$ ($\hat{a}^\dag$) is the annihilation (creation) operator for the cavity field, and $\hat{\sigma}_-=\ket{g}\bra{e}$ ($\hat{\sigma}_+=(\hat{\sigma}_-)^\dag$) is the atomic lowing (raising) operator.

To investigate the SR effect in our system, the first step is to find a bistable region, preparing two metastable states for the occurrence of SR. We first search for the steady-state solutions and study system stability properties in the semiclassical description. By neglecting quantum fluctuations of the field and the atom, we write the classical Langevin equations by replacing quantum operators with classical complex variables $\hat{a}\rightarrow \alpha$, $\hat{\sigma}_-\rightarrow p$, and $\hat{\sigma}_z\rightarrow D_0$:
\begin{eqnarray}\label{alpha}
\dot{\alpha}&=&-\frac{\kappa}{2}\alpha -igp +E_1+E_2 e^{-i\delta t} +\xi\\\label{p}
\dot{p}&=&-\frac{\gamma}{2}p+ig\alpha D_0\\\label{D}
\dot{D}_0&=&-\gamma(D_0+1)-2ig(\alpha p^*-\alpha^*p)
\end{eqnarray}
where we have phenomenologically introduced the cavity decay rate $\kappa$ and the atomic relaxation rate $\gamma$. The stochastic thermal noise $\xi$ satisfies $\aveg{\xi(t)\xi(t')}=2D \delta(t-t')$ with $D$ being the noise strength. In the absence of the weak signal $E_2 e^{-i\delta t}$, we have the steady-state results for the optical field amplitude $\alpha$ and the atomic population inversion $D_0$ by setting the time derivatives in Eqs.(2-4) to zeros:
\begin{eqnarray}\label{bi}
E_1&=&\frac{\kappa\alpha}{2}(\frac{2C}{1+|\alpha|^2/n_0}+1)\\\label{biD}
4E_1^2 D_0&=& n_0 \kappa^2(1-D_0)(1+2CD_0)^2
\end{eqnarray}
where we have defined the cooperation coefficient $C=2g^2/(\kappa\gamma)$ and the saturation photon number $n_0=\gamma^2/(8g^2)$. Under the resonance condition, we can find that $\alpha$ is real and Eqs. (\ref{bi},\ref{biD}) are cubic
equations for $\alpha$ and $D_0$, which in principle have three roots in suitable parameter regimes. Eq. (\ref{bi}) reproduces the familiar expressions for optical absorptive bistability in the JC model \cite{Carmichael,Bergou} and the bistability appears for the cooperation coefficient $C>4$.  Here we choose $C=7.2$ and in Fig. 2 we plot the bistability curves for the optics and atom. We can find that the optical bistability and atomic bistability share exactly same region $E_1 \in [2.15, 2.43]$,  in which $\alpha$ and $D_0$ have three solutions with the red color represents the unstable solution and the black color represents the stable solution. The stability properties were determined by the standard linear analysis method \cite{Strogatz}. Under such parameter setting, the cooperation coefficient $C$ is larger than 4 and the saturation photon number ($n_0=0.3472$) is smaller than unity, which indicates that the system operates at the strong coupling regime but the coupling is not too strong and it is feasible in current technology.

We then include the noise $\xi$ and study stochastic dynamics in the absence of the weak signal. By choosing a driving amplitude in the middle of the bistable region ($E_1=2.25\kappa$), we numerically show the random transitions of the system dynamics between two metastable states activated by the thermal noise $D=0.03\kappa$ in Fig.3 $\textbf{a}$. Clear sharp transitions can be seen and the transitions of the optical field and the atomic population inversion are completely synchronized. The corresponding distributions of two metastable states of the field is shown in Fig.3 $\textbf{b}$ and it exhibits the expected bimodal structure. The distribution of the low amplitude state (L) is a narrow and high peak while the distribution of the high amplitude state (H) is a wider and lower peak. These behaviors can be verified from the potential function.
The effective position variable for the cavity field can be defined as $x=(\alpha+\alpha^*)/2=\alpha$ and the approximate effective potential function can be derived by the relation $\ddot{x}+\frac{\kappa}{2}\dot{x}=-\partial U(x,t)/\partial x$):
\begin{eqnarray}\label{po}
U\left( x\right) & =& U(\alpha)\\\nonumber
 &\approx& \frac{\kappa\gamma}{8}\alpha^2-\frac{\gamma E_1}{2}\alpha+\frac{\gamma^2}{16} \rm{ln}(1+8g^2\alpha^2/\gamma^2)
\end{eqnarray}
In Fig.3$\textbf{c}$ we plot the potential function $U(\alpha)$ of the optical field using the same parameters as Fig. 3$\textbf{a}$ and one can see asymmetric double wells: a wide well at high amplitude and a narrow well at low amplitude. It is well consistent with the distributions in Fig. 3$\textbf{b}$. It is worthy to note that there is correlation between the width of potential wells and the variance of the amplitude fluctuations. The wider the potential well, the larger variance of the dynamics for the corresponding metastable state.
In Fig.3 $\textbf{d}$$(\textbf{e})$ we show the distributions of the residence time $\tau$ for the high (low) amplitude state.  We use the exponential decaying function to fit the data and then we can estimate the average residence times $\bar{\tau_H}=\bar{\tau_L}\approx\bar{\tau}=400 \kappa^{-1}$.
\begin{figure*}[pbt]
  \centering
  \includegraphics[width=6.8in]{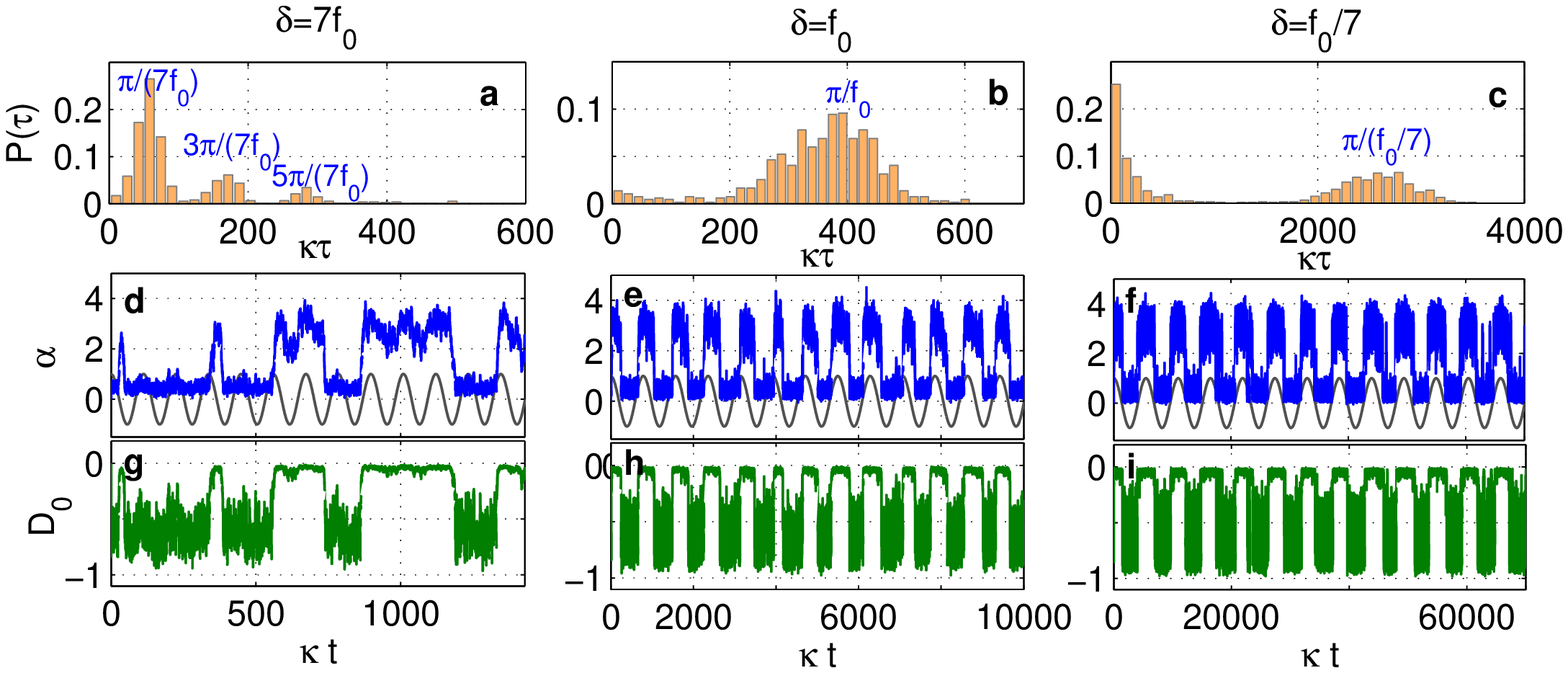}\\
  \caption{Residence time distributions ($\textbf{a}-\textbf{c}$), optical responses ($\textbf{d}-\textbf{f}$) and atomic responses ($\textbf{g}-\textbf{i}$) of the system subject to the weak signal and the noise for three values of the modulation frequency $\delta$ in the semiclassical description: $\delta=7f_0$ for $\textbf{a}$ $\textbf{d}$ $\textbf{g}$, $\delta=f_0$ for $\textbf{b}$ $\textbf{e}$ $\textbf{h}$, and $\delta=f_0/7$ for $\textbf{c}$ $\textbf{f}$ $\textbf{i}$. The parameters are: $\kappa=1$, $\gamma=10\kappa$, $g=6\kappa$, $E_1=2.24\kappa$, $E_2=0.1\kappa$, $D=0.03\kappa$, and $f_0=0.008\kappa$}\label{SRcl}
\end{figure*}

\section{SR phenomena in the semiclassical frame}
In last section we found the bistable region and studied the thermal noise activated random transitions between two metastable states of the system in the absence of the weak modulation signal $E_2 e^{-i\delta t}$. In this section we add this signal to the system and study the SR phenomena using the semiclassical description [Eqs.(2-4)].

Now we have to fix two parameters in order to observe SR: a subthreshold amplitude $E_2$ and a suitable modulation frequency $\delta$ of the signal. The first one is easy to determine by switching off the noise ($D=0$): if the system experiences interwell transitions with this signal, it is a signal over the threshold; else, it is a subthreshold signal. One can choose an amplitude slightly below the threshold. To determine a suitable modulation frequency $\delta$, we recall the average transition time $\bar{\tau}$ obtained in last section. As we know, the matching condition for SR is that the average transition time of the noise induced random transitions equals half of the period of the external signal \cite{Gammaitoni}, that is, $\bar{\tau} =T_{E2}/2$. Then, we can compute the optimal frequency under these parameters as $f_0=2\pi/T_{E2}=\pi/\bar{\tau}\approx 0.008\kappa$.

In Fig. \ref{SRcl}, we plot the residence time distributions of the high amplitude state of the field (the top row) and single trajectories of system responses ($\alpha$ in the middle row and  $D_0$ in the bottom row) in the presence of a subthreshold weak signal $E_2 e^{-i\delta t}$ for three modulation frequencies ($7f_0$, $f_0$, and $f_0/7$). For the left column, the modulation frequency is much higher than the optimal frequency, that is, $\delta=7f_0$. If we merely look at the system responses ($\alpha$ and $D_0$), the dynamics seems random, similar to spontaneous transitions in case of no signal. However, the residence time distribution shows its correlation with the input signal and its distinguishing difference from noise activated spontaneous transitions: there are several peaks with almost constant distance between adjacent two peaks and the locations of peaks are well consistent with the relation $\tau=\frac{T_{E2}}{2}(2n+1)$ with $n=0, 1, 2...$ \cite{Gammaitoni}.

The typical trajectories of system responses to the signal at the optimal frequency $\delta=f_0$ are presented in the middle column. The system dynamics is well synchronized with the dynamics of input signal. Correspondingly, the peak at half periodicity of the signal occupies the majority part of the residence time distribution, which is a signature of the SR effect.

For a frequency lower than the optimal one ($\delta=f_0/7$), as shown in the right column, the periodicity in the system responses remains good but the transitions becomes much noisier. There are two peaks in the residence time distribution: one peak from the noise activated random transitions following the Kramers law and the other locates at the half signal period. Compared to the optimal case (middle column), the proportion of the signal peak occupies much less. Therefore, we have confirmed that only the situation with matched signal and noise leads to best SR effect.

\begin{figure}[pbt]
  \centering
  \includegraphics[width=3.5in]{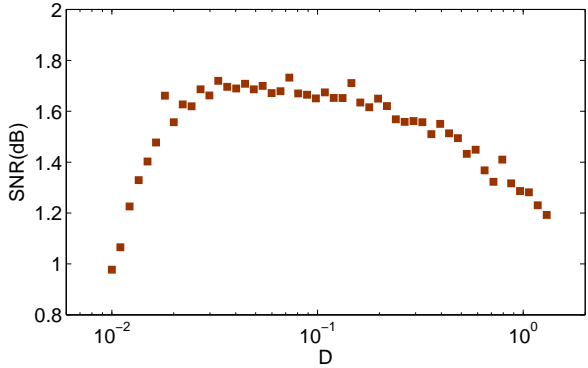}\\
  \caption{ The SNR (dB) as a function of the thermal noise strength $D$ at the modulation frequency $\delta=f_0=0.008\kappa$. Every points are averaged by 10 times. The other parameters are the same as in Fig. \ref{SRcl}.}\label{SNR}
\end{figure}
We then present another feature of SR: a resonant-like peak of the SNR curve. In Fig. \ref{SNR} we plot the SNR in the unit of dB as a function of the thermal noise strength $D$. As expected, the SNR first increases then decreases as the noise $D$ increases and the SNR peaks at a wide range from $0.03\kappa$ to $0.1\kappa$. The SNR (dB) is obtained by evaluating the signal peak hight over the noise background level in the Fourier spectrum. The reason for using logarithm scale for x axis is that the SNR rises rapidly in the low noise range and it drops very slow in the large noise range.

In fact, the optimal values of $D$ and $\delta$ in Fig. 4 are not the only choice to achieve best SR effect with fixed system parameters ($\kappa$,$\gamma$ and $g$). According to the Kramers rate $r_k\propto \rm{exp}(-\Delta V/D)$, the average transition rate of random transitions increases as the noise strength $D$ increases and the required signal frequency for satisfy matching condition is also larger. Therefore, if we increase either one of $D$ and $\delta$, the other one should be increased to certain value correspondingly in order to achieve best SR effect.
\section{Stochastic resonance in the full quantum frame}
\begin{figure*}[pbt]
  \centering
  \includegraphics[width=6.5in]{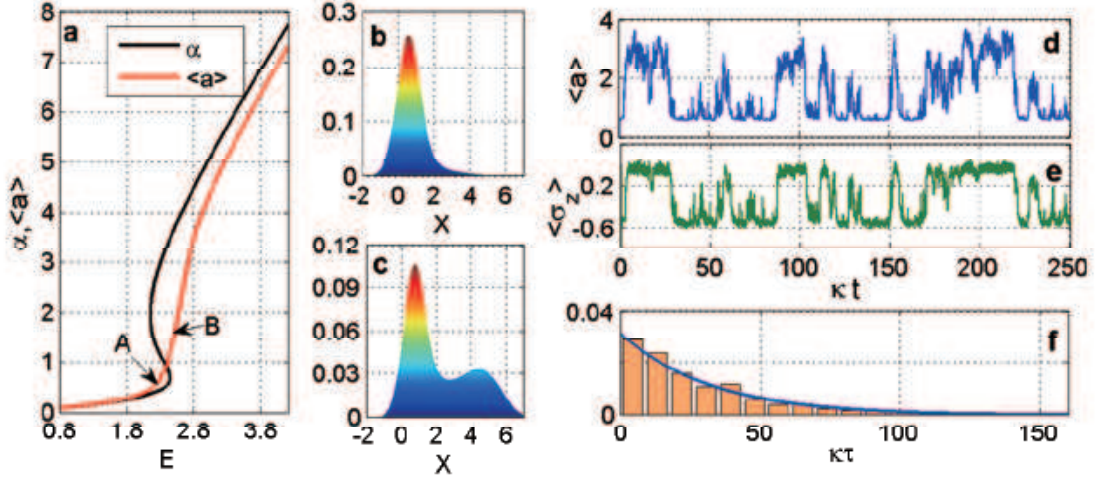}\\
  \caption{$\textbf{a}$ The mean amplitudes of the cavity field in the semiclassical ($\alpha$) and full quantum $\aveg{\hat{a}}$ regimes. The parameters are: $\kappa=1$, $g=6\kappa$, and $\gamma=10\kappa$. $\textbf{b}$ and $\textbf{c}$ are the Wigner function distributions of the optical mode corresponding to point A and point B in $\textbf{a}$. $\textbf{d}$ ($\textbf{e}$) gives a typical trajectory for the optical field amplitude $\aveg{\hat{a}}$ (the atomic population inversion $\aveg{\hat{\sigma}_z}$), obtaining from the SME [Eq.(\ref{SME})]. $\textbf{f}$ shows the residence time distribution of the high amplitude state of the field from a long time evolution ($T=50000\kappa^{-1}$). The exponential fit function is $1/32 \rm{exp}(-\tau/32)$.}\label{qu_Scurve}
\end{figure*}
In preceding sections we found the parameter region for bistability and SR in the semiclassical picture. Now we turn to search for the conditions of SR in the full quantum regime and explore whether quantum fluctuations can induce the occurrence of SR.

First we want to know how much the quantum steady-state results deviate from the semiclassical analyses in the absence of noises. The unconditional dynamics of the system is governed by the unconditional master equation:
\begin{eqnarray}
\label{ME}
\dot{\rho}= i[\rho,\hat{H}] + \mathcal{D}[\sqrt{\kappa}\hat{a}]\rho +\mathcal{D}[\sqrt{\gamma}\hat{\sigma}_-]\rho
\end{eqnarray}
where $\hat{H}$ was given in Eq.(1) and the superoperator $\mathcal{D}$ is defined as $\mathcal{D}[\hat{A}]\rho = \frac{1}{2}(2\hat{A}\rho\hat{A}^\dag-\hat{A}^\dag\hat{A}\rho-\rho\hat{A}^\dag\hat{A})
$.
In Fig.6 \textbf{a} we compare the semiclassical amplitude of the field $\alpha$ with the ensemble averaged amplitude of the cavity field $\aveg{\hat{a}}$ as a function of the input driving $E_1$. We can see that the sharp transition of the quantum curve is not located in the middle of the classical bistability region, instead, it shifts to the larger driving side. This is because that the factorization used in obtaining the semiclassical Langevin equations is not a good approximation for a large coupling coefficient $g\gg \kappa$. To verify this, in Fig. \ref{qu_Scurve}$\textbf{b}$ and $\textbf{c}$ we plot the Wigner function distributions at two different driving strength $E_1=2.25\kappa$ (labeled by point A in $\textbf{a}$) and $E_1=2.55\kappa$ (labeled by point B in $\textbf{a}$). For $E_1=2.25\kappa$, the system is in the region of the semiclassical bistability and it is the parameter used for studying the semiclassical SR in Fig. 4.  However, from $\textbf{b}$ one can see that there is only a single peak in the phase space, which means that at this driving the system is actually monostable. For $E_1=2.55\kappa$ the system operates roughly at the middle point of the quantum curve in $\textbf{a}$ but it is almost at the high amplitude state in terms of the semiclassical curve. At this driving strength the Wigner function distribution exhibits clear double-peak structure. Now we can come to the conclusion that in the full quantum mechanical description one needs stronger driving field to reach the true bistable regime. Therefore, in the following we will choose increased $E_1$ and $E_2$ for investigating quantum SR phenomena.
\begin{figure*}[pbt]
  \centering
  \includegraphics[width=6.5in]{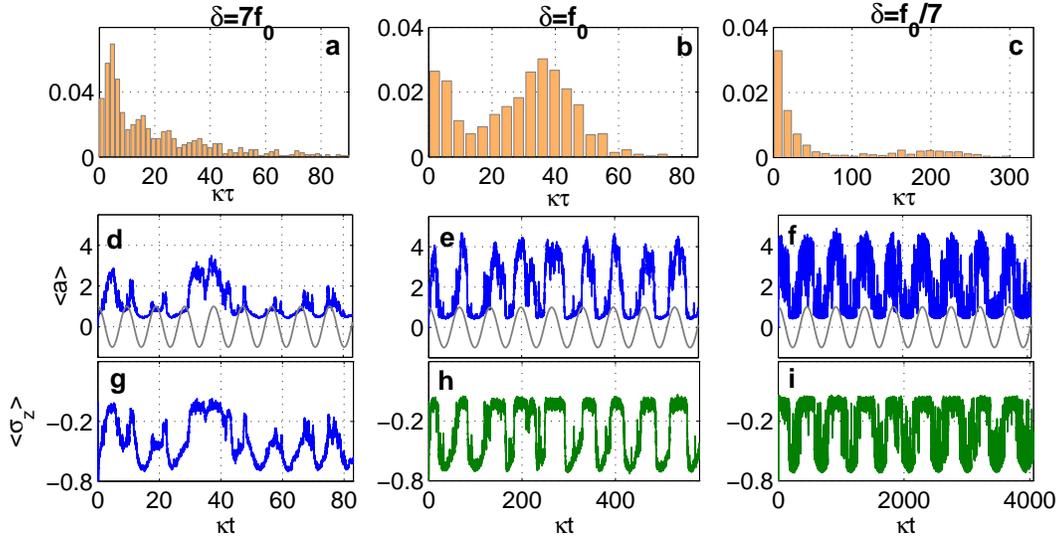}\\
  \caption{Vacuum fluctuations induced SR. $\textbf{a-c}$ the probability distributions of the residence time at the high amplitude state of the field. $\textbf{d-f}$ the responses of the optical mode to the vacuum noise and signal. $\textbf{g-i}$ the responses of the atomic mode to the vacuum noise and signal. The signal modulation frequencies for the left, middle, and right column are $7f_0$, $f_0$, and $f_0/7$ ($f_0=0.095\kappa$), respectively.  The parameters are the same as Fig. 6 except $E_2=0.3\kappa$. }\label{qu_sR}
\end{figure*}
To mimic the realistic model with the inclusion of the quantum noise and signal detection process, we simulate the system dynamics using the quantum trajectory method \cite{Wiseman}. The system dynamics conditioned on the homodyne detection can be described by the stochastic master equation (SME) ($\hbar=1$):
\begin{eqnarray}
\label{SME}
d\rho=dt( i[\rho,\hat{H}] + \mathcal{D}[\sqrt{\kappa}\hat{a}]\rho +\mathcal{D}[\sqrt{\gamma}\hat{\sigma}_-]\rho) +dW(t) \mathcal{H}[\sqrt{\kappa}\hat{a}]\rho
\end{eqnarray}
where $dW$  is the Wiener increments \cite{Gardiner} satisfying $\aveg{dW}=0$ and $\aveg{dW(t)^2}=dt$. The superoperator $\mathcal{H}$ is defined as $\mathcal{H}[\hat{A}]\rho= \hat{A}\rho+\rho\hat{A}^\dag-\rm{Tr}[\hat{A}\rho+\rho\hat{A}^\dag]\rho$.
The corresponding homodyne current is
\begin{eqnarray}
I(t)=\sqrt{\kappa}\aveg{\hat{a}+\hat{a}^\dag}+dW(t)/dt
\end{eqnarray}
Note that in order to highlight the effect of the quantum noise rather than the thermal noise here we consider the reservoir at zero temperature.
Now we want to see whether stochastic transitions can be induced by pure quantum fluctuations at this condition. We choose a driving $E_1=2.55\kappa$ at which the Wigner distribution is bimodal, and we plot the conditional dynamics of the system in Figs. 6 $\textbf{d}$ and $\textbf{e}$. We can see clear quantum jumps in the dynamics and the jumps in optical and atomic modes are exactly simultaneous, similar results can be found in Ref. \cite{Mavrogordatos}. Compared to the semiclassical results, the quantum bistable dynamics has more small spikes and the transitions are less sharp.

To observe SR, we apply the signal field $E_2 e^{-i\delta t}$ to the system. Again, we need to choose a suitable amplitude and a suitable frequency for the signal. In quantum case, we can not judge the over-threshold or subthreshold signal from interwell transitions or intrawell transitions in the absence of the system noise. When noise is off, the system dynamics described by the unconditional master equation [Eq.(\ref{ME})] is an ensemble average of the conditional dynamics and at every time the values of system variables are average values between two metastable states. In this situation, we can judge an over-threshold signal from whether the system dynamics is synchronized to the signal for an arbitrary frequency. A subthreshold signal can only induce good periodic system responses at a suitable frequency and the synchronization will be destroyed especially at a frequency larger than the optimal value. The optimal frequency can be determined using the same procedure as we used in the semiclassical case: obtaining the average transition time from the residence time distribution of noise induced spontaneous transitions and then calculate the optimal external signal frequency using the matching condition of SR. We have shown the distributions of the residence time at the high amplitude state of the field in Fig. 6$\textbf{f}$ and from the fitting data we can obtain the the average transition time $\bar{\tau}\approx 32\kappa^{-1}$ and then we can obtain the approximate optimal signal modulation frequency $f_0=0.095\kappa$.

In Fig.7 we show representative trajectories of system responses ($\aveg{\hat{a}}$  and $\aveg{\hat{\sigma}_z}$) and the corresponding residence time distributions subject to the input signal $E_2 e^{-i\delta t}$ for three values of the modulation frequency $\delta=7f_0, f_0,$ and $f_0/7$. One may find that the SR behaviors in the quantum picture are qualitatively similar to the situation in the semiclassical picture: the system responses are synchronized to the signal best at optimal modulation frequency $f_0$;  a larger frequency $7f_0$ leads to poor periodicity and a smaller frequency $f_0/7$ leads to noisier dynamics. However, there are several differences. Firstly, the optimal frequency $f_0$ is different from the semiclassical case, due to different noise level induced different average transition rate. Secondly, the transitions between metastable state are less sharper and correspondingly the boundaries of peaks in the residence time distributions are more ambiguous compared to the semiclassical results. Thirdly, the required amplitudes of the control field and the signal field are higher in quantum case, due to shift of the quantum bistable region from the semiclassical bistable region.

\section{Conclusion}
We have studied stochastic resonance phenomena in the driven dissipative Jaynes-Cummings model in both semiclassical and full quantum frameworks. Simultaneous occurrence of SRs in the optical and atomic freedoms have been numerically observed. In particular, at zero temperature vacuum fluctuations can drive the spontaneous bistable transitions and the SR effect, in which the input signal is amplified significantly. By comparing quantum SR with semiclassical SR, we find that they are qualitatively similar but the parameter region of quantum SR is shifted from the semiclassical region due to the invalid factorization in obtaining the semiclassical equations in the strong coupling regime. Our results lay theoretical basis for experimental investigating SR in JC model such a fundamental quantum optics system.

\section*{Acknowledgement}
The authors would like to thank Dr Zhenglu Duan for helpful discussions. We gratefully acknowledge financial support from the National Natural Science Foundation of China under Grants No.11504145, No.11364021, No.11664014 and No.11464018, and the Natural Science Foundation of Jiangxi Province under Grants No.20161BAB211013, No.20161BAB201023, and No.20142BAB212004.

\end{document}